\documentstyle[pra,aps,preprint,epsf]{revtex}

\begin{document}
\draft
\title{Biphoton in a Dispersive Medium}
\author{Alejandra Valencia, Maria V. Chekhova$^*$, Alexei Trifonov$^\dagger$, and Yanhua
Shih}
\address{Department of Physics, University of Maryland, Baltimore
County, Baltimore, Maryland 21250} \maketitle
\date{Revised 21 February, 2002}

\begin{abstract}
We report an experimental study of group-velocity dispersion
effect on an entangled two-photon wavepacket, generated via
spontaneous parametric down-conversion and propagating through a
dispersive medium. Even in the case of using CW laser beam for
pump, the biphoton wavepacket and the second-order correlation
function spread significantly. The study  and  understanding of
this phenomenon is of great importance for quantum information
applications, such as quantum communication and distant clock
synchronization.
\end{abstract}

\vspace{1cm} \pacs{PACS Number: 42.50.-p, 42.50.Dv, 03.65.Ta}
Entangled two-photon light has been proved to be useful in quantum
metrology \cite{metrology} and quantum communications \cite{QI}.
New applications have been recently proposed, such as quantum
computing, quantum information processing, and quantum lithography
\cite{lithography}. In all applications, the crucial feature is
the measurement of the correlation of the two-photon entangled
state.  In many cases, the spreading of the biphoton wavepacket
and the second-order Glauber correlation function \cite{Glauber}
becomes a critical issue, especially for these applications in
which the precise timing information is essential, such as
synchronization of distant clocks.

Consider a very simple experiment. A pair of entangled photons is
generated from spontaneous parametric down-conversion(SPDC)
\cite{DNK} and propagates through a dispersive medium. The
dispersive medium could be in one path or both paths. Two
single-photon counting detectors are used for the detection of the
signal and the idler photons, respectively. In most of the
two-photon interferometric experiments, coincidence counting rate
is the only necessary measured quantity:
\begin{eqnarray}\label{0}
R_{c} \sim \int_0^T dt_{1} dt_{2}S(t_{1}-t_{2}-t_{0})
G^{(2)}(t_{1}, r_{1};t_{2}, r_{2}),
\end{eqnarray}
where $G^{(2)}(t_{1}, r_{1}; t_{2}, r_{2})$ is the second-order
Glauber correlation function \cite{Glauber},
$S(t_{1}-t_{2}-t_{0})$ the coincidence window function, which is
usually a rectangular function centered at $t_{0}$, $T$ the data
collection time for the coincidence measurement, $t_{i}$ the
detection time of the $i$-th detector and $r_{i}$ the optical path
to the $i$-th detector. The second-order Glauber correlation
function is defined as
\begin{eqnarray}\label{1}
G^{(2)}(t_{1}, r_{1};t_{2},r_{2})= <E^{(-)}(t_{1},
r_{1})E^{(-)}(t_{2},r_{2})E^{(+)}(t_{2},r_{2})E^{(+)}(t_{1},r_{1})>,
\end{eqnarray}
where $E^{(-)}$ and $E^{(+)}$ are the negative-frequency and the
positive-frequency field operators and ensemble averaging is done
over the quantum state. In the stationary case, $G^{(2)}$ depends
only on $t_1-t_2$. In the case of two-photon state generated via
SPDC, $G^{(2)}$ can be represented as the square modulo of the
two-photon wave function, or biphoton,
\begin{eqnarray}
\psi(t_{1},r_{1};t_{2},r_{2})\equiv\left<0\right|E^{(+)}(t_{2},r_{2})E^{(+)}(t_{1},r_{1})
\left|\Psi\right>,
\end{eqnarray}
where $\left<0\right|$ denotes the vacuum state and
$\left|\Psi\right>$ the two-photon state of SPDC \cite{Rubin}. It
is necessary to point out that in most of the two-photon
interferometry-type measurements, in which the coincidence time
window is much greater than the width of $G^{(2)}$, the temporal
broadening information of the biphoton and the $G^{(2)}$ function
is lost. To measure the temporal correlation function with or
without broadening, one needs to measure $G^{(2)}(t_{1}, r_{1};
t_{2},r_{2})$, but not the time integral of $G^{(2)}(t_{1}, r_{1};
t_{2},r_{2})$.   In the low counting rate cases, a
TAC~(Time-to-Amplitude-Converter)~-~MCA~(Multi-Channel-Analyzer)
system can be used to observe the shape of ${G}^{(2)}$, which
contains the group-velocity broadening information of a biphoton.
Photocurrent pulses from the first and second detectors are sent,
respectively, to the ``start" and ``stop" inputs of a
time-to-amplitude converter (TAC), where the time differences
between ``start" and ``stop" are converted linearly to different
amplitudes of the output pulses. The amplitudes of the output
pulses of the TAC are analyzed by means of a multi-channel
analyzer (MCA). The time resolution of the ``start-stop" method is
mainly determined by the resolution of the detectors, usually in
the order of a few hundreds of picoseconds. The shape of
${G}^{(2)}$ for two-photon light emitted via SPDC can be easily
calculated using Eq.~(\ref{1}), where ensemble averaging should be
done over the state generated in SPDC,
\begin{eqnarray}\label{3}
\left|\Psi\right>=\left| \hbox{vac}
\right>+c\int_{-\infty}^{\infty}d\Omega
F(\Omega)a_i^{\dagger}(\frac{\omega_p}{2}-\Omega)a_s^{\dagger}(\frac{\omega_p}{2}+\Omega)
\left|\hbox{vac}\right>.
\end{eqnarray}
Here the indices $s$ and $i$ denote the signal and idler photons,
respectively; $\omega_p$ is the pump frequency, and
$\Omega\equiv\omega_s-\omega_p/2=\omega_p/2-\omega_i$. The
coefficient $c$ depends on the pump amplitude and the quadratic
susceptibility. We consider the case of cw pump and, for
simplicity, collinear frequency-degenerate phase matching. The
spectral amplitude, $F(\Omega)$, provides all information about
the correlation properties of two-photon light. The spectrum of
signal-idler radiation is $S(\Omega)=|F(\Omega)|^2$. Even if a
dispersive medium of length $z_s$ ($z_i$) is included in the
propagation path of the signal (idler) photon, the first-order
correlation function ${G}^{(1)}$ of the signal (idler) radiation
is given by the Fourier transform of the spectrum only,

\begin{eqnarray}\
{G}^{(1)}(\tau)\sim\int_{-\infty}^{\infty}d\Omega
|F(\Omega)|^2\hbox{cos}(\Omega\tau),
\end{eqnarray}
where
$\tau\equiv[t_{2}-(z_{s,i}/u_{s,i}+r_{2}/c)]-[t_{1}-(z_{s,i}/u_{s,i}+r_{1}/c)]$,
and $u_{i,s}$ denote group velocities of the signal (idler)
photons in the medium; while the second-order correlation function
is calculated to be
\begin{eqnarray}\
{G}^{(2)}(\tau)\sim\left|\int_{-\infty}^{\infty} d\Omega
F(\Omega)e^{i(k_{s}^{\prime\prime}z_{s}+k_{i}^{\prime\prime}z_{i})
\Omega^{2}/2}\cos{(\Omega\tau)}\right|^2,
\end{eqnarray}
where
$\tau\equiv[t_{2}-(z_{i}/u_{i}+r_{2}/c)]-[t_{1}-(z_{s}/u_{s}+r_{1}/c)]$
and we have considered second-order dispersion derivatives
$k_{s}^{\prime\prime}$, $k_{i}^{\prime\prime}$ for the signal and
idler, respectively. In the case of CW pumping, ${G}^{(2)}$
depends only on $\tau$. One can notice immediately that the shapes
of ${G}^{(1)}$ and ${G}^{(2)}$ can be quite different, although
both of them are associated with the spectral amplitude.  It is
clear to see that ${G}^{(2)}$ gets broadened in a transparent
dispersive medium while ${G}^{(1)}$ remains unbroadened. The form
of the spectral amplitude $F(\Omega)$ for SPDC is well-known
\cite{DNK}. For type-II SPDC (and for type-I non-degenerate SPDC),
\begin{eqnarray}\
F_{II} (\Omega)=\frac{\hbox {sin}(DL\Omega/2)}{DL\Omega/2},
\label{4}
\end{eqnarray}
where $L$ is the crystal length and $D$ is the inverse group
velocity difference for signal and idler photons. In the case of
type-I frequency-degenerate SPDC, the shape is
\begin{eqnarray}\
F_{I}(\Omega)=\frac{\hbox{sin}(D''L\Omega^2/2)}{D''L\Omega^2/2},
\label{5}
\end{eqnarray} with $D''$ denoting the second derivative
of the dispersion dependence $k(\omega)$ in the nonlinear medium
for signal and idler photons. The corresponding shapes of the
two-photon amplitude $F(\tau)$ can be obtained from the Fourier
transforms of Eqs.~(\ref{4}) and (\ref{5}). The typical natural
width of ${G}^{(2)}$ for SPDC (without group-velocity broadening)
is in the order of hundreds of femtoseconds, which is impossible
to measure directly due to the relatively low time resolution of
the existing photon counting detectors. It might seem that the
${G}^{(2)}$ shape could be measured by means of the
``anti-correlation "effect \cite{Hong}\cite{Shih}\cite{Steinberg}
(or other two-photon interference effect), which is based on
measuring coincidence counting rate. However, one can show that
the shape of the ``anti-correlation" ``dip" is given not by
${G}^{(2)}$ but by ${G}^{(1)}$ \cite{dip}. The situation is very
similar to that of the classical case: one has to use an
``auto-correlator" to measure the group-velocity broadening of a
laser pulse, instead of an interferometer. Unlike the first-order
correlation function and the spectrum, the unbroadened
second-order correlation function for SPDC radiation is so far
impossible to observe even by using the current state-of-the-art
photon counting detectors. However, after the group-velocity
broadening, for example propagating along a dispersive optical
fibre of certain length, it may be broadened enough to be
measurable. In the present paper, we demonstrate experimentally
how to monitor the broadened ${G}^{(2)}$ by using the TAC-MCA
system, which is quite useful and important in certain types of
two-photon correlation measurements.

Consider the propagation of two-photon light through a dispersive
transparent medium. The spectrum of SPDC and hence, the
first-order correlation function do not change. However, the
second order correlation function ${G}^{(2)}$ does change in this
case.  The situation is analogous to the propagation of short
classical laser pulses through a group-velocity dispersive medium
\cite{Akhmanov}: although the spectrum of a pulse does not change,
its shape changes, being broadened. In the far-field zone, i.e.,
at $z\gg z_{dis}$, where the dispersion length $z_{dis}=\tau_0^2 /
2\pi k''$ and $\tau_0$ is the initial pulse length, the pulse
length varies as $\tau=\tau_0 z/z_{dis}$, and the pulse takes the
shape of the initial spectrum.  A surprising fact in the case of
the entangled two-photon light is that there are no ``pulses",
either broadened or unbroadened, associated with either signal or
idler photons; it is the biphoton and the second-order correlation
function that behave like a ``pulse", which is broadened in the
dispersive medium! In the far-field zone, the shape of the
second-order correlation function ${G}^{(2)}$ is
\begin{eqnarray}\
{G}^{(2)}(\tau,z)\sim |F(\Omega)|^2\large\vert_{\Omega=
\tau/(k_i''z_i+k_s''z_s)}, \label{6}
\end{eqnarray}
i.e., has the same shape as the spectrum of the SPDC. To
demonstrate this behavior experimentally, we used the setup shown
in Fig.~\ref{setup}. A BBO crystal (we used both type-I and
type-II samples cut for degenerate collinear phase matching) was
pumped by an Ar laser at wavelength $457.9nm$. After passing the
SPDC crystal, the pump radiation was cut by a mirror with high
reflection at the pump wavelength and high transmission at the
signal - idler wavelength.  For further blocking of the noise,
RG715 color glass filter was placed behind the mirror. The angular
selection of the signal-idler radiation was done by a $2mm$
pinhole placed at a distance of $1.5m$ from the crystal. After
passing the aperture, the signal and idler radiation was fed into
a single-mode polarization-maintaining fibre of length $2m$ or
$500m$, for different experimental demonstration. The slow and
fast axes of the fibre were aligned according to the polarization
directions of the ordinary and extraordinary rays in the BBO
crystal. After the fibre, the output beam was split by a
polarizing beamsplitter and sent to two single-photon counting
modules (Perkin\&Elmer SPCM-AQR-14). In the case of type-II BBO,
the PBS separated signal and idler photons. In the case of type-I,
a half-wave plate after the fibre was used to rotate the
polarization by $45^\circ$; this way, the output part of the setup
worked as a Hanbury Brown-Twiss interferometer with a $50\%$
beamsplitter. For a better blocking of the noise, color glass
filters RG830 were inserted in front of the detectors. The
photocurrent pulses from the first and second detectors were sent,
respectively, to the ``start" and ``stop" inputs of a TAC. Since
the counting rate of the detectors did not exceed $10^5$ counts
per second, the distribution at the MCA output corresponded to the
${G}^{(2)}(\tau)$ function \cite{MW}. To find the time resolution
of the setup, we first measured the MCA distribution for the case
of $5$ mm type-II BBO crystal without the use of optical fibre
(Fig.~\ref{fig2a})~ \cite{MCA}. According to this measurement, the
resolution of the setup is found to be $700ps$. For a $3mm$
type-II BBO, passing the signal and idler radiation through a
$500m$ optical fibre does not spread much the observed MCA
distribution. Indeed, the width increase of the second-order
correlation function after the fiber should be $50$ ps in this
case, which is much less than the resolution. However, for a
$400\mu$ type-II BBO, the initial ${G}^{(2)}$ is a rectangle with
the width of $60f$s.  The calculated value of $z_{dis}$ is in this
case $1.8cm$. After the fibre, the width of the ${G}^{(2)}$ should
be about $3ns$, which is measurable by means of our TAC-MCA setup.
The resulting distribution is plotted in Fig.~\ref{fig2b}. The
shape of the ${G}^{(2)}$ becomes similar to the initial spectrum
of type-II SPDC shown in Eq.(6), with distinct side lobes. The
solid line in the plot is a fit using Eq.(6), the only fitting
parameter being $k''$ for the fibre. The parameter $D$ was
calculated from the Selmeier equations for BBO; for wavelength
$916nm$ and cutting angle $36.6^\circ$ of collinear degenerate
phase matching, $D=1.5(ps/cm)$. For $k''$ of the fibre, we
obtained a value of $3.2\cdot10^{-28}(s^2 / cm)$, which is very
close to the values typically observed for single-mode fibres
(see, for instance,
 \cite{fibre}). The SPDC spectrum for $3.4mm$ type-I BBO is broader
than the spectrum for $0.4mm$ type-II crystal. Correspondingly,
the ${G}^{(2)}$ broadening by a $500m$ fibre (Fig.~\ref{fig2c}) is
even greater than that for Fig.~\ref{fig2b}.  The shape resembles
the spectrum for type-I SPDC.
%the absence of side lobes is probably
%caused by absorption at the signal-idler wavelength.
The theoretical calculation (solid line) was done with the same
value of $k''$ as shown in Fig.~\ref{fig2b}. As $D''$, we took the
figure $D''=5.9\cdot10^{-28}(s^2 /cm)$, which is calculated from
the results of ${G}^{(1)}$ measurement in type-I BBO \cite{fake}.

Consider now what happens if narrow-band filters are inserted in
front of the detectors $D_{1}$ and $D_{2}$. In this case, the
filters would cut the central part of the SPDC spectrum. The
effect is the same as if $F(\Omega)$ were narrowed or the initial
${G}^{(2)}(\tau)$ were broadened. Hence, the insertion of filters
would reduce the broadening effect of the fibre. In agreement with
this, we observed almost no spreading for a $500$ m optical fiber
if $10$ nm interference spectral filters were inserted in front of
$D_{1}$ and $D_{2}$, see Fig.~\ref{fig2d}.

Finally, two important conclusions should be formulated.

1. Although both first-order and second-order correlation
properties of two-photon light generated via SPDC are initially
determined by the spectral amplitude function, they behave
differently after the propagation in a dispersive medium. It is
necessary to find an effective method of monitoring the broadening
of the second-order correlation function $G^{(2)}(\tau)$ for
certain two-photon entanglement applications.

2. Passing a biphoton through a dispersive medium one can affect
the shape of the biphoton wavepacket. To be precise, one can at
least spread the second order correlation function as we have
shown. To our belief, the reverse process of compressing the width
of the two-photon wavepacket is also possible, of course only
within the reach of a Fourier-transform limit. A dispersive medium
with the reverse coefficient may be applied in this case
\cite{Franson}. A special type of optical fibre, diffraction
grating or a prism pair can be used for this purpose. The
situation is almost identical to the manipulation of the shape of
a short laser pulse \cite{compression}; however, it is the
manipulation of the biphoton wavepacket but not the signal or the
idler. This effect is of high importance for the applications of
two-photon and multi-photon states in quantum communication and
quantum information processing.

We are grateful to M.Rubin for illuminating discussions. This work
was supported in part by NSF, ONR and NRO. MVC and AST also
acknowledge support from INTAS 01-2122.

$^*$Permanent address: Dept. of Physics, Moscow State University,
Moscow, Russia.

$^\dagger$Permanent address: MagiQ Technologies Inc., 11 Ward St.
Suite 300, Somerville, MA 02143. Also affiliated with A.F. Ioffe
Physical Technical Institute RAS, 26 Polytekhnicheskaja, 194021
St. Petersburg, RUSSIA

\begin{figure}[hbt]
\caption{The experimental setup.} \label{setup}
\end{figure}

\begin{figure}[hbt]
\caption{The MCA distribution for the case of a $5mm$ type-II BBO
crystal and no fibre. The distribution is basically determined by
the time resolution of the detectors.} \label{fig2a}
\end{figure}

\begin{figure}[hbt]
\caption{The MCA distribution for the case of an $0.4mm$ type-II
BBO crystal and a $500m$ fibre. The shape is similar to the
spectrum for type-II SPDC.} \label{fig2b}
\end{figure}

\begin{figure}[hbt]
\caption{The MCA distribution for the case of a $3.4mm$ type-I BBO
crystal and a $500m$ fibre. The shape is similar to the spectrum
of type-I SPDC.} \label{fig2c}
\end{figure}

\begin{figure}[hbt]
\caption{The MCA distribution for the case of $0.4mm$ type-II BBO
crystal, $500$ m fibre, and $10nm$ filters inserted in front of
the detectors. As a result, the $G^{(2)}(\tau)$ function is not
spread, and the width of the distribution remains the same as for
the case of a short fibre.} \label{fig2d}
\end{figure}

\newpage
\centerline{\epsfxsize=3in \epsffile{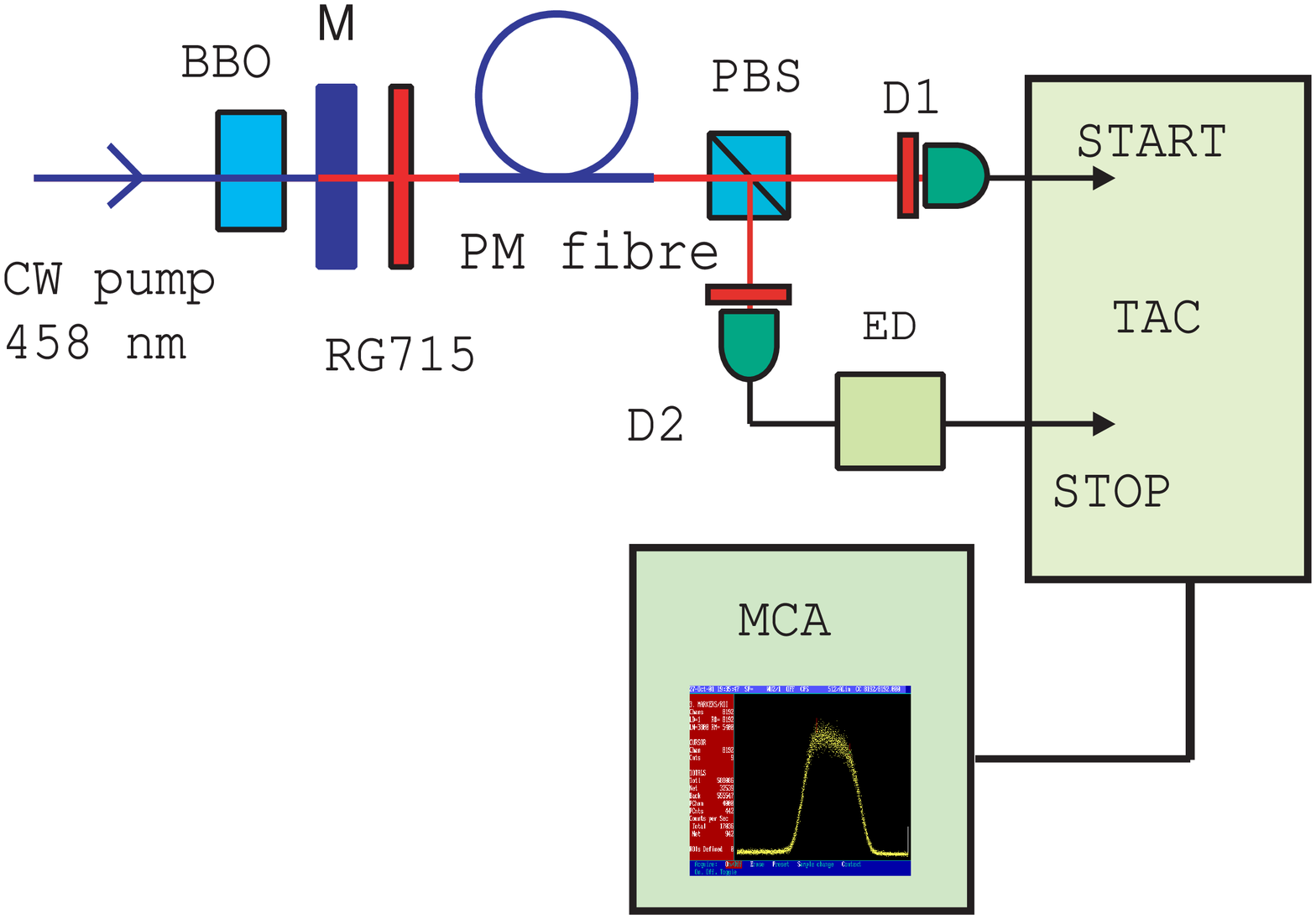}} \vspace{1cm}
Figure \ref{setup}.  Alejandra Valencia, Maria V. Chekhova, Alexei
Trifonov, and Yanhua Shih.

\newpage
\centerline{\epsfxsize=3in \epsffile{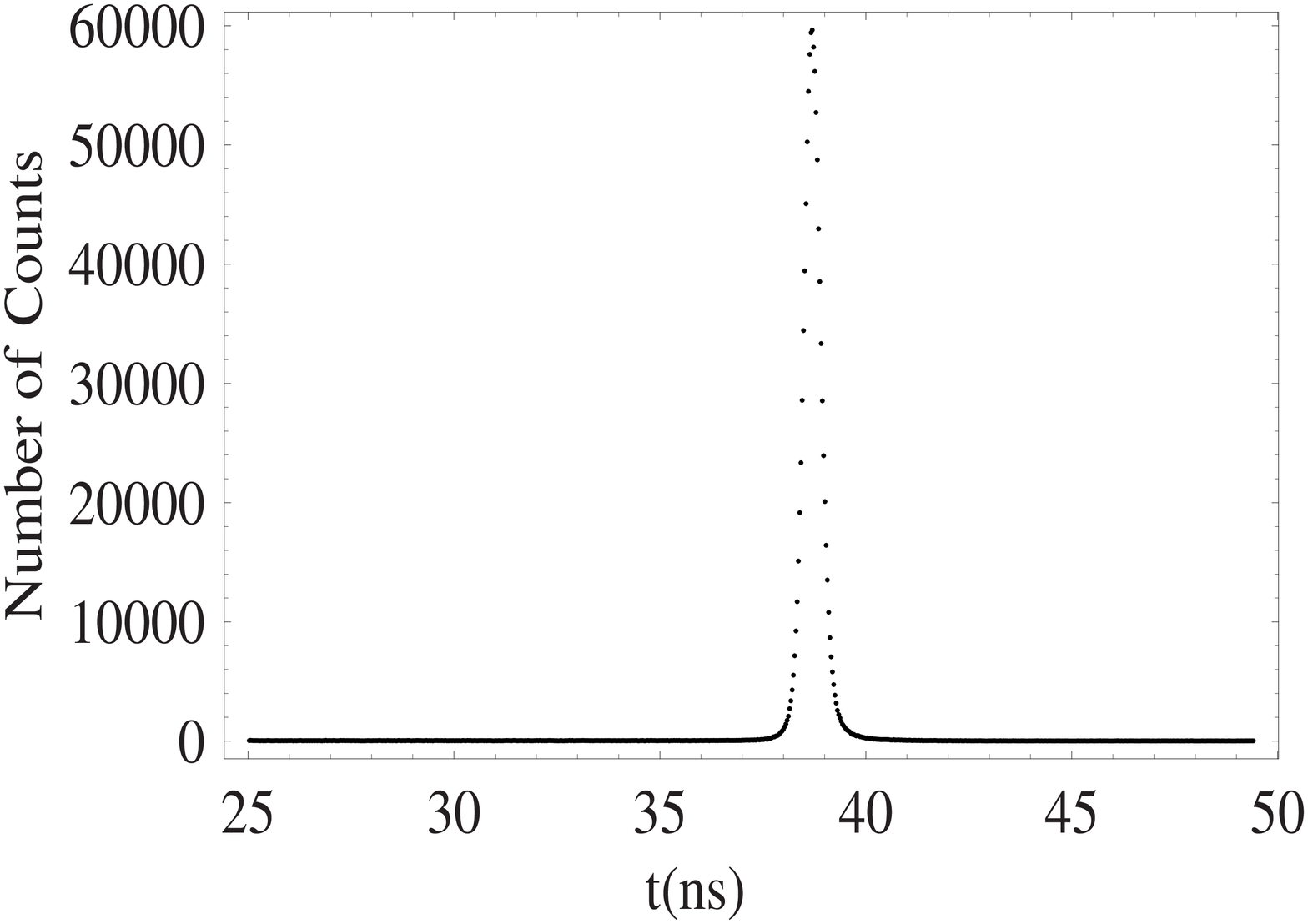}} \vspace{1cm}
Figure \ref{fig2a}.  Alejandra Valencia, Maria V. Chekhova, Alexei
Trifonov, and Yanhua Shih.

\newpage
\centerline{\epsfxsize=3in \epsffile{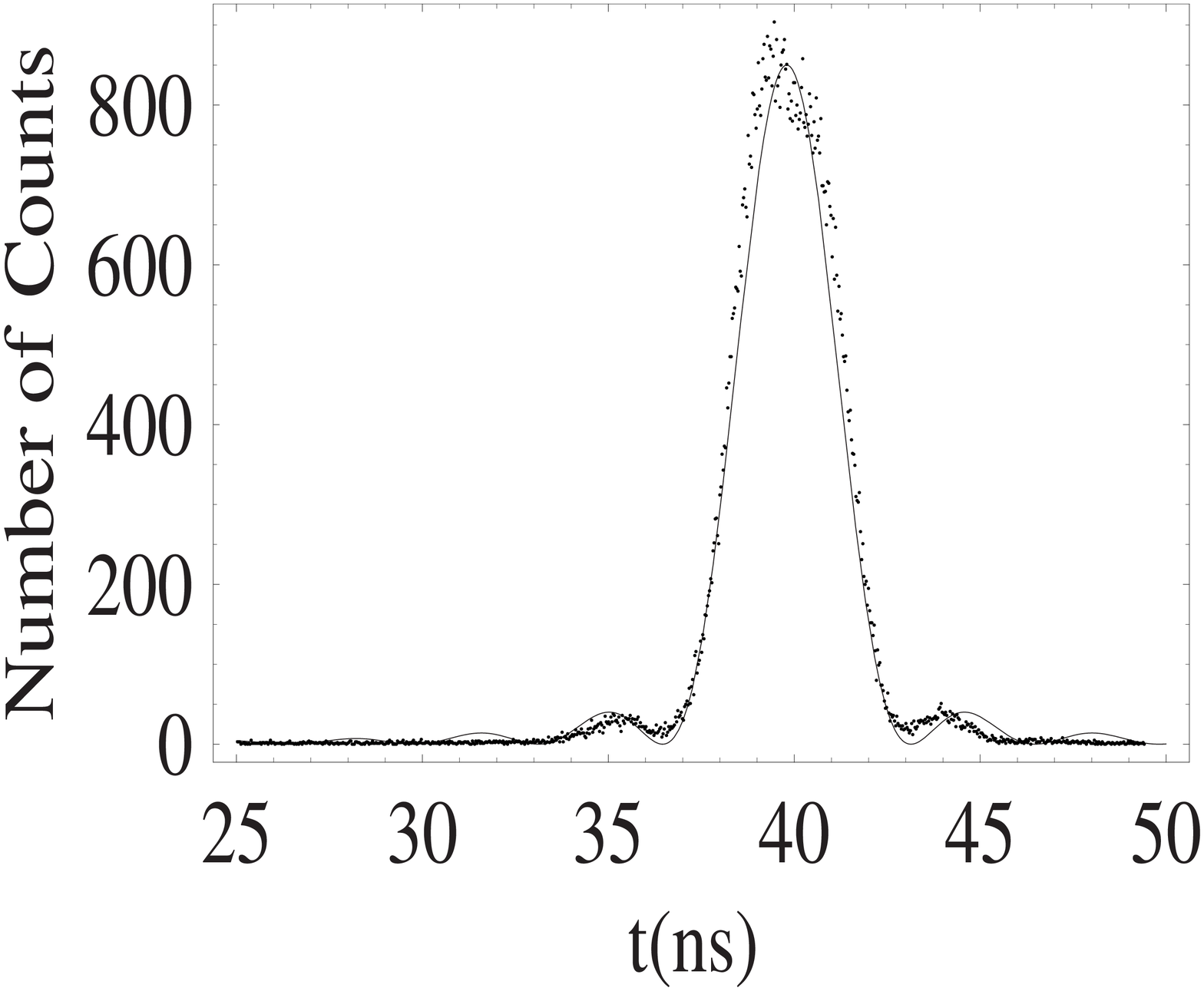}} \vspace{1cm}
Figure \ref{fig2b}.  Alejandra Valencia, Maria V. Chekhova, Alexei
Trifonov, and Yanhua Shih.

\newpage
\centerline{\epsfxsize=3in \epsffile{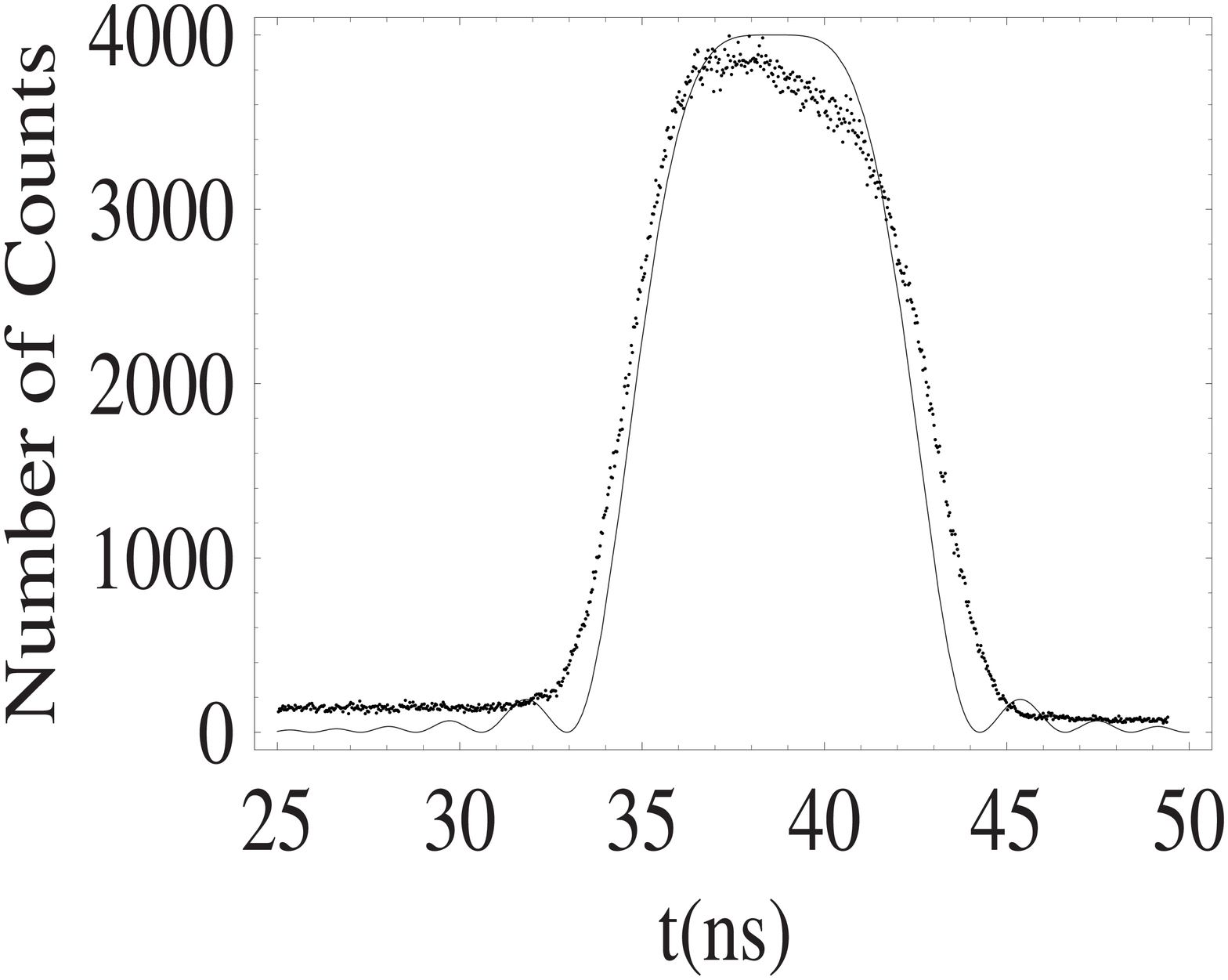}} \vspace{1cm}
Figure \ref{fig2c}.  Alejandra Valencia, Maria V. Chekhova, Alexei
Trifonov, and Yanhua Shih.

\newpage
\centerline{\epsfxsize=3in \epsffile{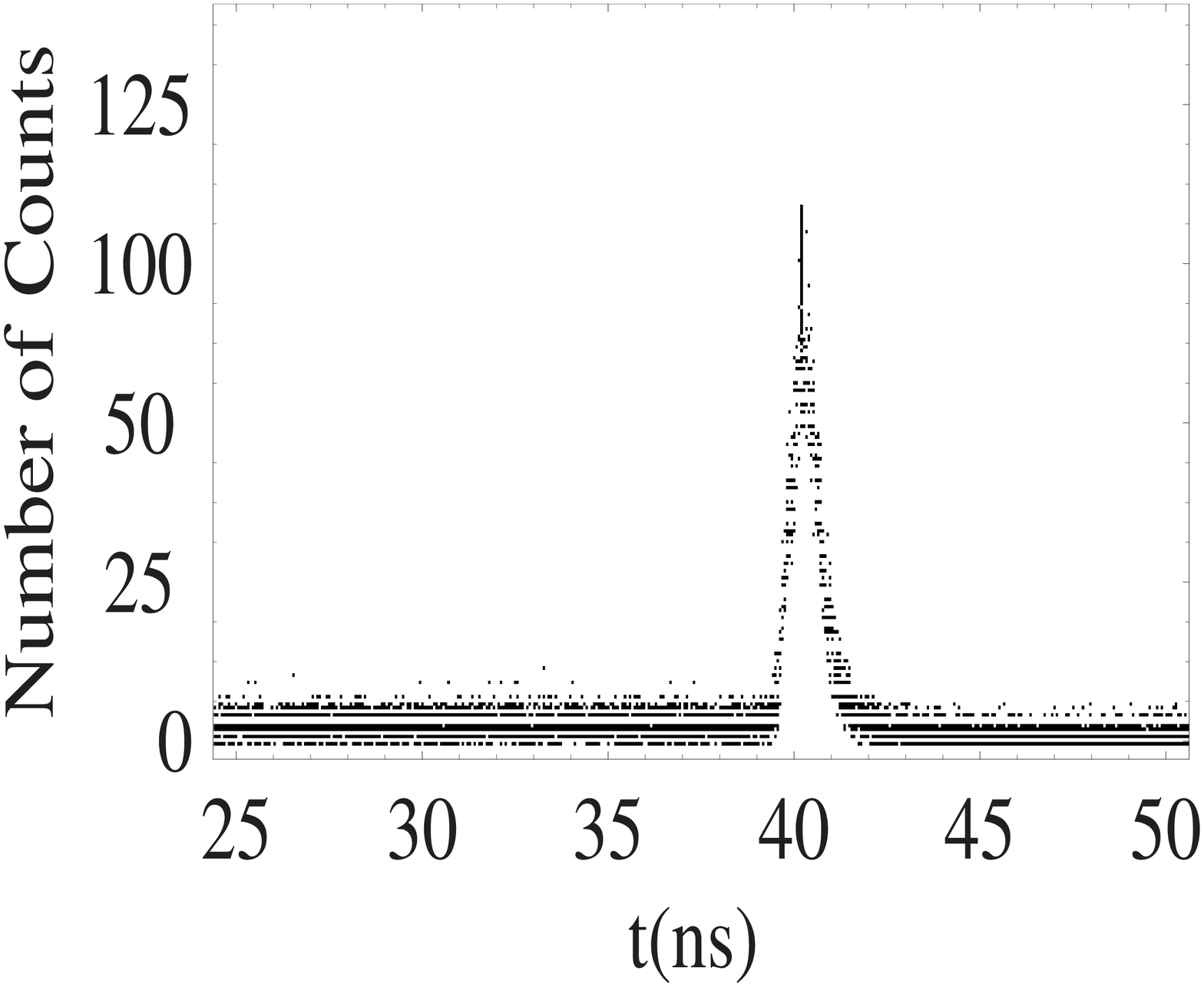}} \vspace{1cm}
Figure \ref{fig2d}.  Alejandra Valencia, Maria V. Chekhova, Alexei
Trifonov, and Yanhua Shih.


\begin{references}
\bibitem{MCA} For better accuracy, most of MCA distributions were
averaged over $10$ channels.
\end{references}
\end{document}